%% file: ieeeGitHubMicrosoft.tex
\documentclass[journal,twocolumn]{IEEEtran}

\ifCLASSINFOpdf
\else
\fi

\usepackage[utf8]{inputenc}
\usepackage{booktabs}
\usepackage{amsmath}
\usepackage{amssymb}
\usepackage{paralist}
\usepackage{multirow}
\usepackage{xspace}
\usepackage{color}
\usepackage{xcolor}
\usepackage{tcolorbox}
\usepackage{ifthen}
\usepackage{url}
\usepackage{fancybox}
\usepackage{enumitem}
\usepackage{listings}
\usepackage{ifthen}
\usepackage[tight,footnotesize]{subfigure}
\usepackage{graphicx}
\usepackage{amsmath}
\usepackage[linesnumbered]{algorithm2e}
\usepackage{tablefootnote}
\input{macros}

\graphicspath{{pics/}}

\definecolor{codegreen}{rgb}{0,0.6,0}
\definecolor{codegray}{rgb}{0.5,0.5,0.5}
\definecolor{codepurple}{rgb}{0.58,0,0.82}
\definecolor{backcolour}{rgb}{0.95,0.95,0.92}

\lstdefinestyle{mystyle}{
	backgroundcolor=\color{backcolour},   
	commentstyle=\color{codegreen},
	keywordstyle=\color{magenta},
	numberstyle=\tiny\color{codegray},
	stringstyle=\color{codepurple},
	basicstyle=\footnotesize,
	breakatwhitespace=false,         
	breaklines=true,                 
	captionpos=b,                    
	keepspaces=true,                 
	numbers=left,                    
	numbersep=2pt,                  
	showspaces=false,                
	showstringspaces=false,
	showtabs=false,                  
	tabsize=2
}

\usepackage[english]{babel}

\DeclareGraphicsExtensions{.pdf,.jpeg,.png}
\begin{document}

\newtheorem{theorem}{Definition}[section]
	
	\newcommand{\eg}{e.g.,}
	\newcommand{\ie}{i.e.,}

	\newcommand{\one}{small library}
	
	\newcommand{\ones}{small libraries}
	
	\newcommand{\tool}{DepChainsJS}
	\newcommand{\twoandtwoplus}{2nd and 3rd}

	\newcommand{\RqOne}{\textit{How prevalent are micro-packages?}\xspace}
	
	\newcommand{\RqTwo}{What is the impact of micro-package dependencies?\xspace}
	
	\newcommand{\RqThree}{\emph{What is the developer usage cost of a \texttt{npm} package?}\xspace}
	
	\renewcommand{\lstlistingname}{Listing}

\newcommand{\boxedtext}[1]{\fbox{\scriptsize\bfseries\textsf{#1}}}
\newcommand{\nota}[2]{
	\boxedtext{#1}
		{\small$\blacktriangleright$\emph{\textsl{#2}}$\blacktriangleleft$}
}
\newcommand{\todo}[1]{{\color{red}\nota{TODO}{#1}}}

\title{FLOSS != GitHub: A Case Study of Linux/BSD Perceptions from Microsoft's Acquisition of GitHub}

\author{
\IEEEauthorblockN{
Raula Gaikovina Kula\IEEEauthorrefmark{1},
Hideki Hata\IEEEauthorrefmark{1},
Kenichi Matsumoto\IEEEauthorrefmark{1}}\\
\IEEEauthorblockA{\IEEEauthorrefmark{1}Nara Institute of Science and Technology, Japan\\
\{raula-k, hata, matumoto\}@is.naist.jp}
}

\maketitle

\begin{abstract}
In 2018, the software industry giants Microsoft made a move into the Open Source world by completing the acquisition of mega Open Source platform, GitHub.
This acquisition was not without controversy, as it is well-known that the free software communities includes not only the ability to use software freely, but also the libre nature in Open Source Software.
In this study, our aim is to explore these perceptions in FLOSS developers.
We conducted a survey that covered traditional FLOSS source Linux, and BSD communities and received 246 developer responses.
The results of the survey confirm that the free community did trigger some communities to move away from GitHub and raised discussions into free and open software on the GitHub platform.
The study reminds us that although GitHub is influential and trendy, it does not representative all FLOSS communities. 
\end{abstract}

%
\IEEEpeerreviewmaketitle

\section{Introduction}
\label{sec:intro}
Microsoft's GitHub is now reported to support a community where more than 40 million people learn, share, and work together to build software.
The git-based platform has grown to around 100 million repositories and has released some of the world’s most influential technologies, such as React Native by Facebook, Tensorflow by Google, and Swift by Apple just to name a few. 
GitHub is based on social coding, which introduces online collaborations to attract and sustain developers to a project \cite{Kula:2019}.
In June of 2019, Tidelift and The New Stack jointly fielded a survey of professional software developers to show that most developers (84\%) view themselves as active contributors to open-source projects~\cite{survey2019}.

On October 26th 2018, Microsoft announced on its official blog that it had completed its acquisition of GitHub~\cite{blog2018}. 
In a blog Nat Friedman outlined two principles for GitHub~\cite{nat2018}:
\begin{enumerate}

    \item \textit{{GitHub will operate independently (from Microsoft) as a community, platform, and business.}} This means that GitHub will retain its developer-first values, distinctive spirit, and open extensibility. 
    \item \textit{{GitHub will retain its product philosophy.}} We love GitHub because of the deep care and thoughtfulness that goes into every facet of the developer’s experience. 
\end{enumerate}
With these statements, GitHub confirms its intention to retain and grow its base of Open Source developers.

From a traditional point of view, the free software community represents a different aspect of the open source software community.
It is described as a campaign for computer users' freedom and the open source camp that only focus on practical benefits of software.
To be fair to all camps, we use the term “FLOSS,” meaning “Free/Libre and Open Source Software"~\cite{floss}.
Although the free software community has had its share of disagreements with Microsoft~\cite{barr2013,wikipedia,windows8,fsf2018}, the only reported negative opinion of free software community has different attitudes towards GitHub is the idea of `forking' so far, as it it is considered as a danger to FLOSS development~\cite{10.1007/s10664-016-9436-6}.

In this paper, we report on how external events such as acquisition of the open source platform by a closed source organization triggers a FLOSS developers such the Linux/ BSD Free Software communities.


\begin{table*}
\centering
\caption{Targeted Linux Distributions and BSD Communities}
\label{tab:demographics}
\begin{tabular}{@{}ll@{}}
\toprule
 Channel & Communities \\ \midrule
 Forums & Mint, Manjaro,Debian, Solus, Antergos, openSUSE, MX Linux, Zorin, Arch, ReactOS, Lite, Puppy \\
 &FreeBSD, SparkyLinux, Slackware, Xubuntu, Devuan, Bodhi,  Gentoo, Kubuntu, Sabayon, KNOPPIX, 4MLinux, Tiny Core \\
&ClearOS, GhostBSD, NixOS, Ubuntu Studio, NuTyX, wattOS,  LibreELEC, Trisquel, siduction, Porteus, Elive, Scientific \\
&Parabola, Maui, Zenwalk, BunsenLabs, Void, Artix,  Salix, Pardus, FreeNAS, Pinguy, NAS4Free, IPFire \\
&OpenMediaVault, pfSense, Fatdog64, Neptune, SUSE, VyOS,  MiniNo, Arya, Runtu, Peach OSI, SalentOS, Zevenet \\
&3CX, NethServer, Wifislax, ArchStrike, Porteus Kiosk, Funtoo,  KXStudio, Freespire, OviOS, Haiku, Pearl, Karoshi \\
&MINIX, Untangle, LinuxBBQ, Refracta, BigLinux, HardenedBSD, PrimTux, EasyNAS, MidnightBSD, Toutou, TurnKey, DietPi \\
&KANOTIX, Cucumber, Linspire, AsteriskNOW, RISC, CloudReady, Rebellin, RaspBSD, Springdale, Securepoint, PLD, SME Server \\
&Swift, TalkingArch, NexentaStor, SMS, Ulteo, Volumio,  SuliX, Webconverger, DRBL, Dragora, UBports, Liquid Lemur \\
&AIO, SuperGamer, Namib, Obarun \\\midrule
Mailing List & Debian, Fedora, LXLE, ROSA, DragonFly, Calculate, OpenMandriva, IPFire, SliTaz, NetBSD, Uruk, CRUX\\
&heads, Debian Edu, Endian, OSGeo, LuninuX, Whonix\\
&APODIO, Rocks Cluster, Clear, Lunar, Frugalware, GoboLinux\\
&MirOS, Super Grub2, Bio-Linux, GuixSD, Rescatux, gNewSense\\
&Exherbo, Thinstation, Vine, BSDRP, OLPC, T2\\
&Grml, Swecha\\
 \bottomrule
\end{tabular}%
\end{table*}

\begin{table*}
\caption{Questions that were asked to the Respondents}
\label{tab:questionnaire}
\resizebox{\textwidth}{!}{%
\begin{tabular}{@{}ll@{}}
\toprule
 & Survey Questions to be answered using a Likert Scale Ranking \\ \midrule
Respondents & 1. Do you think it would be a good idea to move the  project away from GitHub?  \\
that remain on GitHub & 2. If the project will be moved away from GitHub to another platform, how much \\
&additional effort will be required to get accustomed to the new platform? \\
& 3. If you have any specific comments, please feel free to add here (Optional) \\\midrule
Respondents  & 1. How much do you think this decision to move away from GitHub was related to the acquisition?\\
that moved away  & 2. Do you agree with the decision to move the distribution away from GitHub?\\
from GitHub & 3. Does moving away from GitHub affects your contributions to this project? \\
& 5. How much additional effort will be required to get accustomed to the new platform? \\
& 4. If you have any specific comments, please feel free to add here (Optional) \\\midrule
Respondents & 1. Do you think it would be a good idea to move the project to GitHub?\\
that do not use GitHub  & 2. If the distribution/kernel will be moved to GitHub, how much additional \\
& effort will be required to get accustomed to the new platform?\\
& 3. Apart from Linux and BSD contributions, have you had personal experiences with using GitHub? \\
& 4. If you have any specific comments, please feel free to add here (Optional) \\\midrule
Open-end feedback&  1. If you have any generic comments, please feel free to add here (Optional).\\\bottomrule
\end{tabular}%
}
\end{table*}

\section{Target Subjects and Survey Design}
\label{sec:study}

Table \ref{tab:demographics} shows a listing of the targeted subjects for the study. 
It is important to note that our survey results do not represent the perceptions of all FLOSS developers. 
Instead is only contains a targeted subset of contributors of Linux distributions/kernel and BSD families to infer the opinion of traditional FLOSS development communities.
We distributed our survey via two lesser intrusive communication channels:
\textit{i. \textit{Forums.}} Our attempt to use the forums was to follow developer communication channels. In this case, we posted in the more generic and random channels.
\textit{ii. \textit{Mailing Lists.}} In the absence of a forum, we reverted to a mailing list. 
Since the mailing lists are curated, we had to request the mediating curator for permission before posting the survey. 

Table~\ref{tab:questionnaire} outlines the structure of the questionnaire we designed our online form.
The survey was broken into three main sections, that were based on the respondents activities after the acquisition and are listed below:
\begin{itemize}
    \item \textit{Remaining on GitHub} This section would be filled out by respondents that had GitHub projects and would continue to use GitHub after news of the acquisition. 
    \item \textit{Moved away from GitHub} This section would be filled out by respondents that moved away from GitHub after news of the acquisition. 
    \item \textit{Never used GitHub} This section would be filled out by respondents that have never used GitHub.
\end{itemize}
Finally, we asked all respondents to provide additional open-ended feedback on the acquisition.
The intention was to later classify whether or not the feedback was positive, neutral or negative.

\begin{table}
\centering
\caption{Demographics and Perspective of Linux or BSD Distribution respondents. Respondents can choose more than one demographic.}
\label{tab:stats}
\begin{tabular}{@{}lr@{}}
\toprule
\textbf{ Demographic (multiple choice allowed) }& \textbf{\# of Respondents} \\ \midrule
 Casual contributors & 149\\
 Core contributors & 64 \\
 Project manager/team leader & 47 \\
 Others & 16 \\\midrule

\midrule
\textbf{ Perspective }&  \\ \midrule
 Remain on GitHub & 138 (56\%)\\
 Moved Away from GitHub& 75 (31\%) \\
 Do not use GitHub & 33 (13\%) \\\midrule
 TOTAL & 246 (100\%)\\
 \bottomrule
\end{tabular}%
\end{table}


\subsection{Participant Demographics}
Table \ref{tab:stats} depicts the total respondents to our survey, totalling to 246 total responses.
Considering that we used developer communication channels for our survey, it was interesting that the responses were mainly filled out by many causal contributors (149). 
This was followed by core contributors (64) and then the project manager/leader (47).
Before we proceeded, we asked participants on their impressions of GitHub.
We found that 63\% of the respondents were fans of GitHub.
A majority of the respondents either agreed (46\%) or had not opinion (20\%) whether \textit{`GitHub appeals with access to over 27 million users'}.
Likewise many agreed (58\%) that GitHub has a useful set of functions for developers.
However, developers were not ready to openly compare GitHub when compared to similar platforms, with 59\% of developers not agreeing or having no opinion on whether GitHub is the superior platform.

We report that over half of respondents (55\%) claim that the acquisition would be detrimental to their GitHub projects. 
Furthermore, most respondents negatively responded to the possibility of an expansion of Free and Open source contributors (74\%), and about half of the respondents (45\%) did not think the acquisition would bring improvements, reliability, and services with the platform.

\section{Survey Findings}
\label{sec:findings}
Based on the questionnaire, we first analysed all answers according the perspective and then conducted a deeper analysis of the open-ended optional comments.

\subsection{Feedback based on Perspective}

As shown in Table \ref{tab:stats}, we found that over half of our respondents did not move away from GitHub (138, 56\%), while a third of the respondents already had their projects move away from GitHub (75, 31\%).
Finally, there was a smaller number of respondents that did not host any of their projects on GitHub (33, 13\%).

For developers that are currently contributing to GitHub, 33\% of respondents thought it was a bad idea for their projects to move.
In terms of the additional effort required to move, there was mixed responses, with no clear majority.

For the developers that had any of their projects migrate away, the majority completely agreed (65\%) with the decision to move and were sure that the acquisition was the key motivation (59\%).
Interestingly, when asked about how the move would affect their contributions, developers were less forthcoming with 53\% either neutral or having no opinion.
Still developers were more optimistic with the move, with a majority of 36\% confident that no additional effort would be required to get accustomed to the new platform.

For the developers that never contributed to any GitHub projects, 80\% thought that moving to GitHub was not a good idea.
In terms of the effort required, a significant number of respondents (29\%) had no opinion, while 48\% reported that it would take them much effort to get accustomed.
That being said, a majority (65\%) of these developers had other open source of personal projects on GitHub.

\begin{tcolorbox}
\textbf{Takeaway 1:} Some of the responded developers of the Linux and BSD distribution 
had left GitHub or had not used GitHub, although the majority continued using GitHub.
Developers that stayed with GitHub did not cite that additional effort is not a reason why they remained with GitHub.
\end{tcolorbox}

\subsection{Open-end Feedback}
We also collected all open-end comments (optional) and manually categorized the reasons based on their polarity of their sentiments (i.e., positive, neutral or negative).
Out of the 70 responses, we were removed 9 responses that were not relevant, leaving 61 sentences.

\paragraph{\textbf{Positive Responses}}
We find 3 respondents that were in favor of the acquisition, stating that 
\begin{quote}
    \textit{`Microsoft brings money, money means stable and reliable hosting of our source code'}
\end{quote}
As has been reported in previous studies, innovative services are the key to sustainable projects~\cite{7166083}.
Other sentiments includes the statement that \textit{`It's useful for mirrors in order to make it easy for drive-by contributors'} and that it is convenient, stating that \textit{`I don't see anything wrong with buying Github by Microsoft. For example, we use Skype so far and we will continue to use it until it suits our needs, or until we do something relatively convenient for us.'}

\paragraph{\textbf{Neutral Responses}}
For the  28 respondents, they were not as concerned of the acquisition as much, citing other alternative platforms such as GitLab or that \textit{`It's easy to migrate to other git hosting platforms, but that's not needed right now for me.'}
From this point of view, respondents did cite the benefits of using GitHub because of its huge userbase,
\begin{quote}
    \textit{`The problem is, if too much devs move to other platforms, then the greatest benefit of GitHib (easy cross-contribution and linking between different projects) will be lost. Thus follow the herd ;).'}
\end{quote}
Another respondent expressed that the strong support \textit{`Github has strong support for SaaS integration (CI/CD services in particular), but these services are provided by third parties. Github's competitors .. provide CI/CD functionality as part of their packages which would probably make me decide to use them instead of Github if I had to set up a new user account right now."} and yet another \textit{`Imho GitLab is superior in parts of functionality but is still lacking the userbase.'}

Other respondents cited that GiHub was not their primary source, thus did not see any immediate threat. For example, one respondent stated that \textit{`FreeBSD doesn't use GitHub as its primary source anyway'} and that it was too early to see any affect, \textit{`Practically speaking, nothing has changed yet. We are not even talking about predictions, but about ``gut feelings'''}.
As mentioned by a Debian developer:
\begin{quote}
    \textit{`The Debian project had always been in favour of not using services running on infrastructure that is not managed by Debian. Impact to the project has been minimal as we had already our own infrastructure (Alioth recently migrated to salsa.debian.org, our own gitlab infrastructure). Debian is used to handling upstream developers using many different forges (Sourceforge, Savannah, GitHub, Gitlab) and the impact of upstream moving from GitHub to GitSalsa has not been significant.'}
\end{quote}

This sentiment can be seen in other responses that state that \textit{`The ownership of GitHub is not relevant at all.'}.
Overall, there is divided feelings, even within the teams:
\begin{quote}
    \textit{`Many team members are not concerned about the use of Github, and the convenience of the features on Github are useful. Visibility of the project (example Mapserver) is very important to some team leaders, and there is no replacement. We at OSGeo dot org have already built our own git server, and it is popular in our group.. the content of the projects is perfectly mirrored, but the tickets and PR history does not port. Some people have found tools to migrate away from Github, but it has not happened in general. The opinions about this are very divided, with strong feelings on different sides.'}
\end{quote}

In fact, respondents were quick to remark that GitHub is propriety.
\begin{quote}
    \textit{`The problem is NOT GitHUB-Microsoft deal but more general, a small example: in the past we (nearly) all use usenet. There are TONS of newserver around the world, anyone can host another if he or she want. There are plenty of groups some fun, some about sport, some about work, some to help others, some to discuss new ideas, ... today we have FEW non-integrated proprietary platforms (from Fb to StackExchange to Reddit to 4Chan, YC, ... all of them are POOR replacements of ancient ng. All of them are proprietary and depend on a single company.'}
\end{quote}

However, being propriety itself had mixed feelings, with one respondent stating that:
\textit{`This has always been a proprietary service using open source technologies. The owner doesn't really matter to me as long as they don't mess with code or visibility of projects for business purposes'}.

\paragraph{\textbf{Negative Responses}}
We found 30 negative responses, which expressed a distrust of a closed source to merge with an open source.
For example,
\begin{quote}
    \textit{`A Platform to develop OpenSource software cannot be controlled by King of closed source software. It's a contradiction.'}
\end{quote}
Similar responses where expressed by respondents, even if there was no grounds. 
For example, 
\begin{quote}
    \textit{`It's quite hard for me to trust Microsoft, they're probably still planning to EEE stuff - even a "harmless" tiger raised in captivity will still bite'}
\end{quote}
Ideas of freedom, in terms of both legal and that it could have an effect on the quality of the code. 
Such concerns arised with a respondent that stated that \textit{`.. the ability to change the foundations and the fundamentals of your project by using terms and conditions will result in a poorer product and more people worried about legal implications instead of writing code.'}. Another respondent was concerned with the idea of a monoculture, stating that
\begin{quote}
    \textit{`Monocultures are to be avoided, GitHub should simply be one of a number of similar projects to be used by software projects.'}
\end{quote}

This is supported by another respondent that \textit{`"GitHub is a centralised, proprietary platform and always has been'}.
Another respondent was abit less direct, stating that \textit{`I prefer free and open tools'}.
Another yet, stated how it impedes competitive software (sw), \textit{`MS will impede development of competitive sw to MS products'}.

Finally, there were statements that directly referenced the freedom of open source in their response, stating that:
\begin{quote}
    \textit{`Today and not since today our freedom is getting smaller and smaller from the hardware (UEFI (in)secure-boot etc) to services (how many people instead of have a local maildir sync-ed, for instance, only live on webmails?) to the society in general'}
\end{quote}

\begin{tcolorbox}
\textbf{Takeaway 2:} We found that the theme of the free-nature of Open Source, especially in regards to GitHub Open Source projects was raised during the discussion of the acquisition.
\end{tcolorbox}

\section{A Year Follow-up }
We contacted a subset of the developers on November 12th 2019 to confirm the results and survey their stance, a year after the acquisition and first survey.
The main goal was to confirm our findings by asking the single question:
\textit{After one year, has your perception on Microsoft's acquisition of GitHub changed?}
Out of the 16 emails sent out to the respondents, we received 4 responses.

We found that all respondents had unchanged sentiments towards the acquisition, with stronger conviction on the violation of freedom.
One interesting response we found that GitHub should not be a reflection of all Open Source:
\begin{quote}
    \textit{`All development of open source is happening on Github - therefore Github = Open Source and that I have heard not once but many times and I have also heard Git = Github'}
\end{quote}

Furthermore, the social impact of GitHub on the Open Source community is a concern for a free community developer.

\begin{quote}
    \textit{`What many open source developers see in Github and many other services is a free lunch which they are so it’s hard to argue about that.  It’s a free lunch that encourages you to market the free lunch and that you are almost an outcast if you aren’t. 
    “Hey why aren’t you putting the github link on your page? Don’t you know all open source happens on GitHub?”'}
\end{quote}

\begin{tcolorbox}
\textbf{Takeaway 3:} The acquisition provides a reminder to acknowledge the existence of FLOSS Communities that reside outside the GitHub Platform.
\end{tcolorbox}

\section{Closing Remarks}
We understand and acknowledge that our survey may only represent a small fraction of the free community of traditional Open Source software, but it is a good indicator that the freedom community is still active.
Looking forward, we recognize that although GitHub hosts influential and exciting part of Open Source, it is a reminder that GitHub is not always a representative FLOSS. 


\end{document}

%% file: macros.tex
{\begin{center}\vspace{0mm}\noindent\begin{Sbox}\begin{minipage}{0.95\columnwidth}}%
{\end{minipage}\end{Sbox}\fbox{\TheSbox}\end{center}\vspace{0mm}}